# An Extensive Analytical Approach on Human Resources using Random Forest Algorithm

[1]Swarajya lakshmi v papineni, [2] A.Mallikarjuna Reddy, [3] Sudeepti yarlagadda ,[4]Snigdha Yarlagadda, [5] Haritha Akkineni

[1] Professor, Department of IT,Prasad V Potluri Siddhartha Institute of Technology, vijayawada, AP, India
[2] Assistant Professor, Department of CSE, Anurag University
[3]Freelance HR Consultant,Hyderabad,telangana State, India
[4] Technology analyst, Infosys limited,Bangalore,Karnataka, India
[5] Associate Professor, Department of IT,Prasad V Potluri Siddhartha Institute of Technology,Vijayawada, AP, India

[1] papinenivsl@gmail.com,[2] mallikarjunreddycse@cvsr.ac.in,[3]ysudeepti@gmail.com

**Abstract:** *The current job survey shows that most software employees are planning to change their job role due to high pay for recent jobs such as data scientists, business analysts and artificial intelligence fields. The survey also indicated that work life imbalances, low pay, uneven shifts and many other factors also make employees think about changing their work life. In this paper, for an efficient organisation of the company in terms of human resources, the proposed system designed a model with the help of a random forest algorithm by considering different employee parameters. This helps the HR department retain the employee by identifying gaps and helping the organisation to run smoothly with a good employee retention ratio. This combination of HR and data science can help the productivity, collaboration and well-being of employees of the organisation. It also helps to develop strategies that have an impact on the performance of employees in terms of external and social factors.*

**Introduction:** The proposed system aims to find the employees who are seeking for a new job.The dataset contains 14 features. The features are described as shown in the **table 1**.

### Table 1: Dataset Description& Analysis

| S.No | Name | Description | Data Type | Possible Values | Number of Missing values |
|---|---|---|---|---|---|
| 1 | enrollee_id | Unique ID for enrollee | int64 | A continuous value | 0 |
| 2 | City | City code | Object | city_n, where n represents unique value for city number | 0 |
| 3 | citydevelopmentindex | Developement index of the city | float64 | A continuous value | 0 |
| 4 | Gender | Gender of enrolee | Object | Male, Female and Other | 4508 |
| 5 | relevent_experience | Relevent experience of enrolee | Object | Has relevent experience, No relevent experience | 0 |
| 6 | enrol_university | University type they have enrolled | Object | no_enrollment, Part time course, Full time course | 386 |
| 7 | education_level | Education level of enrolee | Object | Primary School, Graduate, Masters, High School,Phd | 460 |
| 8 | major_discipline | Education major discipline of enrolee | Object | STEM, Business Degree, Arts, Humanities, No Major, Other | 2813 |
| 9 | Experience | Enrolee total experience in years | Object | <1 to >20 | 65 |

| 10 | c_size | No of employees working currently in the organization | Object | <10 to 5000-9999 | 5938 |
| 11 | company_type | Type of current employer | Object | Pvt Ltd , Funded Startup, Early Stage Startup ,Other ,Public Sector, NGO | 6140 |
| 12 | Lastnewjob | Job role difference compared to the previous work profile | Object | Never, 1, 2, 3, 4, >4 | 423 |
| 13 | training_hours | training hours completed | int64 | A continuous value | 0 |
| 14 | Target | Status for looking of new job | float64 | 0: not looking change of job 1: looking for change of job | 0 |

The training dataset contains 19158 records and test dataset contains 2129 records.

The dataset is obtained from the Kaggle[1] open source. The dataset contains heterogeneous data and is good for binary classification.All the machine learning algorithms are classified into 2 types. They are:

(i) Supervised Machine Learning [2]: In this approach, it defines a function that maps features as shown in below equation i, to the predicted variables. The main advantage of this approach is it has labelled data and it trains the data until it achieves the good accuracy values. The supervised algorithms are either deals with regression process or classification process. In the recent years, most of the classification algorithms are working on the ensemble mechanisms. During the training phase, the input values are cross mapped with output values and are trained to predict the correct labels for the test data. In this approach, all the class labels are dependent on the features of the data. Depending on the value of the class label, the model decides either to apply classifier or regressor. If the class label is continuous parameter it applies regression otherwise it applies classifier.

**Output=f(input) – (i)**

The supervised algorithms always predict the class labels based on the previous experiences it has and it also uses these previous experiences to optimize the performance metrics. The only limitation with supervised learning is it suffers from computational time. In this paper, the decision on the classification type plays a key role, since the model have various types of classifiers like linear classifiers, Kernel estimators, and Tree classifiers

A) Linear Classifier: It uses a linear function to identify the class label of the record. These combine the feature vectors with weights to define the linear function, they also defines the boundaries to obtain the separable space to classify the objects. The best algorithms under linear classifiers are support vector machine, naïve bayesian.

Sometimes logistic regression is also considered as "Linear classifier".

B) Kernel estimators: These algorithms are sort of non-parametric values to find the probability of the features. In this algorithm, based on the samples and conditions, some inferences are drawn. The relationship between the feature and class label are represented using "probability distribution function". To estimate the density value, system can use histograms, radial basis function, and bandwidth.

C) Tree Classifiers: In these algorithms, the training of the data occurs based on the simple rules that are generated by the model. There are three nodes available in this tree, root node which contains all the samples of the training data. Internal node, the recursively distributed records of the data based on the attribute values. Leaf nodes contain the predicted class label. To divide the data into subsets system use two step statistical approaches known as information gain and gini index.

(ii) Unsupervised Machine Learning: In this approach, the values belong to the same group are clustered [3] but all these are not labelled outcomes. These algorithms widely used in image processing areas and where reduction of dimensions plays an important role. The sub classification deals with clustering, dimensionality reduction and association mining. In this type of approach, all the algorithms identify the hidden patterns and discover the important aspects of the feature. The accuracy values of this approach are not as so good as supervised algorithms. The best case study for this approach is credit card fraud detection, where it tries to find the outliers using clustering technique to detect the unofficial transactions.

Even though it is time consuming process it doesn't require any manual labelling of data. It is a perfect tool for designing the dimensionality reduction because it clearly understands the data from the scratch. The probability concepts decide the relation between the similar data, which is a powerful statistical approach to analyze the data. There are two types of unsupervised machine learning. They are:

A) Dimensionality Reduction: The computation time increases with the increase of the attributes. The principal component analysis performs this task by transforming the data into smaller subsets and applies the necessary techniques. The variance and covariance foe understanding their importance by their presence in the dataset.

B) Association Mining:It helps in finding the hidden patterns. With the help of support and confidence, the association mining generates the rules to find the relation between feature, A and feature, B. The popular algorithms are apriori, FP-tree growth, Eclat and etc.,

**Related Works:**

I Setiawan [4,5] et al, designed a Employee attrition analysis using logistic regression. This system considers the dataset that describes the number of employees, leaving organization every year. The data analysis is performed on the various attributes like in_time, out_time, satisfaction level, number of working hours, number of ideal hours and etc. After performing the analysis, it has obtained nearly 75% accuracy.

V. Kakulapati[6,7] et al, proposed a predictive analytics model for HR. The model uses unsupervised K-Means clustering algorithm to decide the number of promotions to be given in each department and number of new employees to be hired for every department. The model randomly chooses some centric points and calculates the Euclidean distance. Based on the distance, the features are allocated to the nearest centric points. The model assumes k value as iterations and for k iterations it updates the cluster values to which it belongs to the features.

Dana Pessach [8,9] et al, developed a analytical framework using decision tree to predict the new employees, who are suitable for job position. The model contains two main components. They are: local prediction for hiring new employee based on the highlights of the performance. Global optimization focuses on the maximization of general objectives i.e., every employee is mapped to the roles he/she can fit. Later, the patterns are extracted and sub optimal procedures are applied to find the best suitable role for the employee. In global component, mathematical programming and considers various parameters that represents the candidates, qualifications and positions they held.

IshaanBallal [10,11] et al, proposed a supervised ML classification approach for predicting the people, who are leaving the job. The proposed system founded the features which played a key role in determining the features that impact the reasons for the employees' termination. They adapted random forest algorithm to build the model and achieved a good accuracy and it compared with advanced machine learning algorithms.

Sabina-Cristiana Necula [12-16] defined a data science architecture that can analyze and derive the data for finding the right employee with good technical skills. The semantic web technologies extract the data that are specified in the resume and various skills are identified using the supervised machine learning algorithms. Based on the work experience, education and skills the ontology's in semantic web mining constructs the structured data that are generated from different vocabularies of the resumes in the database. Now, it is the responsibility of the RDF to convert the data from csv and categorize them as classes, sub-classes and properties. The SPARQL tool queries the data and extracts the features. Finally, decision tree classifier generates the rules to predict the class label.

SandeepYadav [17,18] suggested some data mining techniques for prediction of employee attrition. The model contains 12 features that are related to the professional domain. The model has pre-processed the data using the common mechanisms and transformed the values so that they can fit the evaluation. The model has analyzed the correlation matrix and found that only 10 features are sufficient to predict the model. The model has compared five classification algorithms to decide the good accuracy performer. From the studies it has identified that decision tree algorithm performs best with an accuracy of 98.17%.

**Data Visualization& Analysis:** This section specifies about the exploratory data analysis, which represents the inferential statistics and also summarizes data from various aspects. The **figure 1** represents the categorical data distribution with respect to class label. Data visualisation offers easy-to-use and attractive visualization features and resources to help the users to visualize their data in a simple and meaningful way. Analysts can interpret ideas and emerging patterns using the pictorial representation of data sets. Making sense of the quintillion bytes of data generated every day is impossible without Data Proliferation, which involves data visualization.

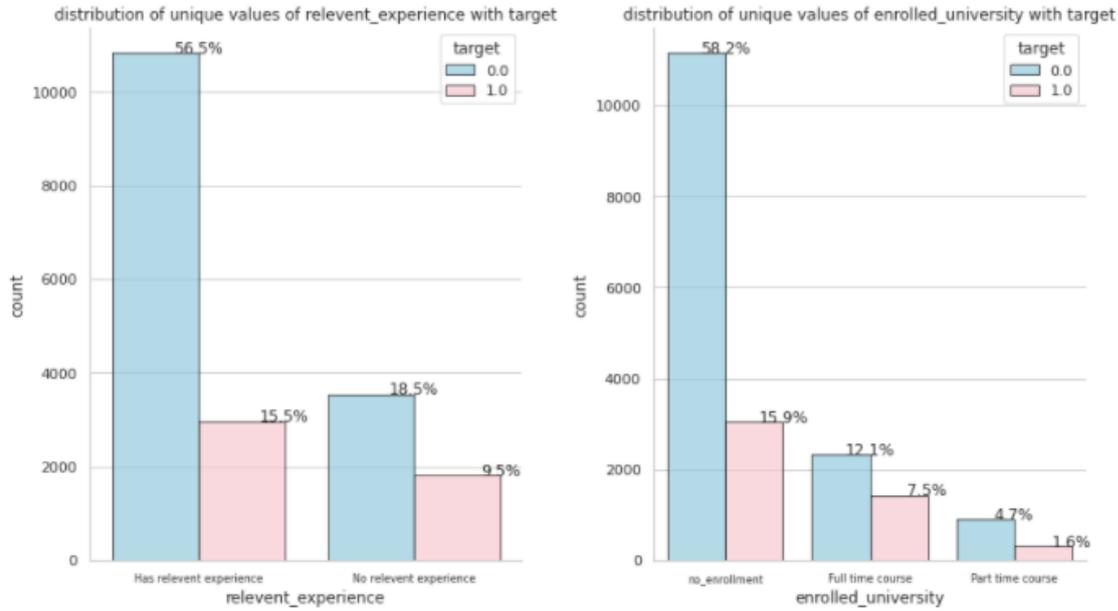

**Figure 1: Categorical Data Distribution over target (Class Label)**

Major Interpretations:

a. number of male employees are greater than number of female employees

b. among the employees who want to change the job are having relevant experience

c. most of employees have not enrolled in any university

d. From the education details system can observe that, a huge number of graduates and a very few PhD holders want to change their job role

e. All most all the employees are from stem background.

Similarly, the numerical data distribution over class label is represented in **figure 2.**

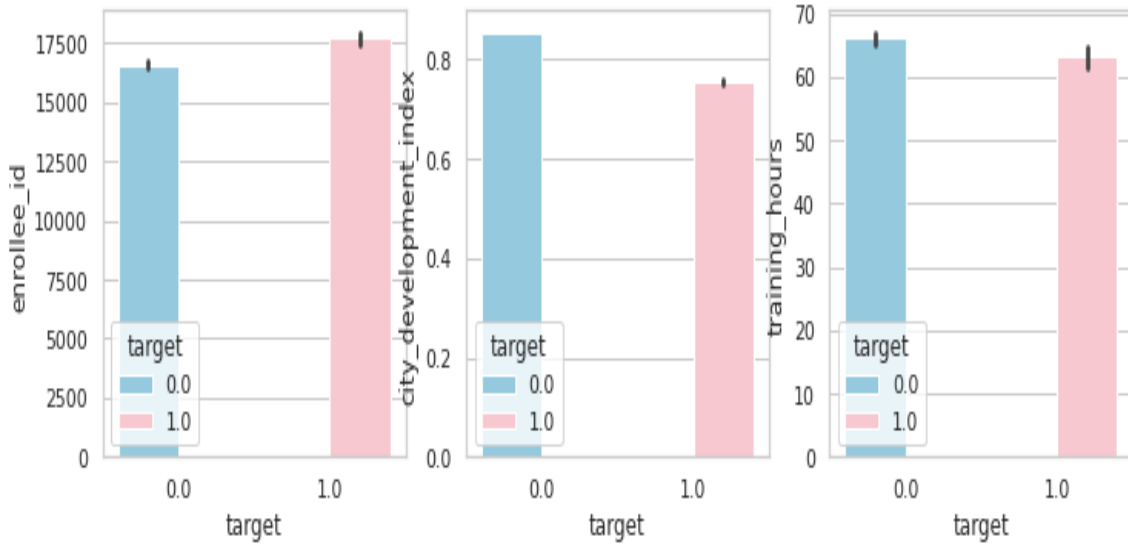

**Figure 2: Numerical Data Distribution**

The correlation matrix between the numerical values and target value of the dataset is shown in the **table 2.**

**Table 2: Correlation of Numerical Features with the Target**

|  | Enrollee_id | City_development_index | Training_hours | target |
|---|---|---|---|---|
| Target | 0.05 | -0.34 | -0.002 | 1.00 |

Interpretations:

a. Training hours and target features are independent

b. Cities development index and target are dependent variables and more over, if the city development index is high then the people who wants to change the job is high.

c. Enrollee id has no impact on the target

**Proposed System:**

The previous studies has implemented various traditional and latest machine learning algorithms and obtained an accuracy of 77%. To improve this accuracy, this paper aims to implement random forest algorithm with cross fold validation. The block diagram of the proposed system is as shown in the **figure 3.**

a. Data Cleaning: The dataset contains many missing fields and this in turn impact the overall performance of the model. To ensure the efficiency, the proposed system used median strategy to fill the missing values of the numerical attributes and mode strategy to fill the categorical values.

b. Data Conversion: Any machine learning model cannot work with textual data. The categorical data are converted into numerical values using label encoding [8] and one hot encoding. The label encoding mechanism first arranges all the unique values in the ascending order of alphabets and assigns the value from 0 to k-1, (where k represents count of distinct values of a particular attribute) as shown in the **table 3.**

**Table 3: Label Encoding Mechanism for major_discipline attribute**

| Categorical Value | Numerical Value |
| --- | --- |
| STEM | 0 |
| Business Degree | 1 |
| Arts | 2 |
| Humanities | 3 |
| No Major | 4 |
| Other | 5 |

The label encoding is not efficient when system have more number of unique values, because with the increase of unique values n value will also increase, this can be handled by one hot encoding, which constructs a matrix with binary values for all the possible values.

c. Data Transformation: Dataset contains heterogeneous data either it can be difference in unit values or it can be range values. The transformation process converts the data into uniform values. This process makes the model to treat all the features with equal priority. In the proposed system, it uses MaxAbsScaler, which takes the maximum value of a particular column and divides the every value with that maximum value. The equation is shown below as **equation 1.**

$$f_i = \frac{f_i}{maximum\ value} \quad - (1)$$

Random Forest Classifier with CV: All the machine learning algorithms now days are updating their values based on the hyper parameter turning. Random Forest algorithm [9] doesn't need any turning process. It constructs a forest of trees based on bagging methods. The random forest finds the best node among the random subsets generated and one more advantage of random forest classifier is it automatically finds the best features based on the impurity index calculated during the generation of trees.In this model, the Leave-One-Out cross validation [10] is implemented to configures the model based on the single hyper parameter K and one subset is considered for test and remaining are considered for training. It has high computational cost still it give robust cost. The main advantage of this model is it considers every row for cross validating, so that the model gets enough training and predicts more accurately. The dataset in this model is medium size, so the usage of this LOOCV will not have any adverse effect on the model. In random forest classifier, with the increase in the number of forest trees, the robustness of the model increases. To generate more trees, the model has selected this LOOCV. The Random forest algorithm is good at handling missing values, categorical values and even it doesn't get suffer from overfitting problem, which is a common problem in many of the ensemble bagging and boosting algorithms. It always considers the majority voted target values to perform the best split at each iteration.

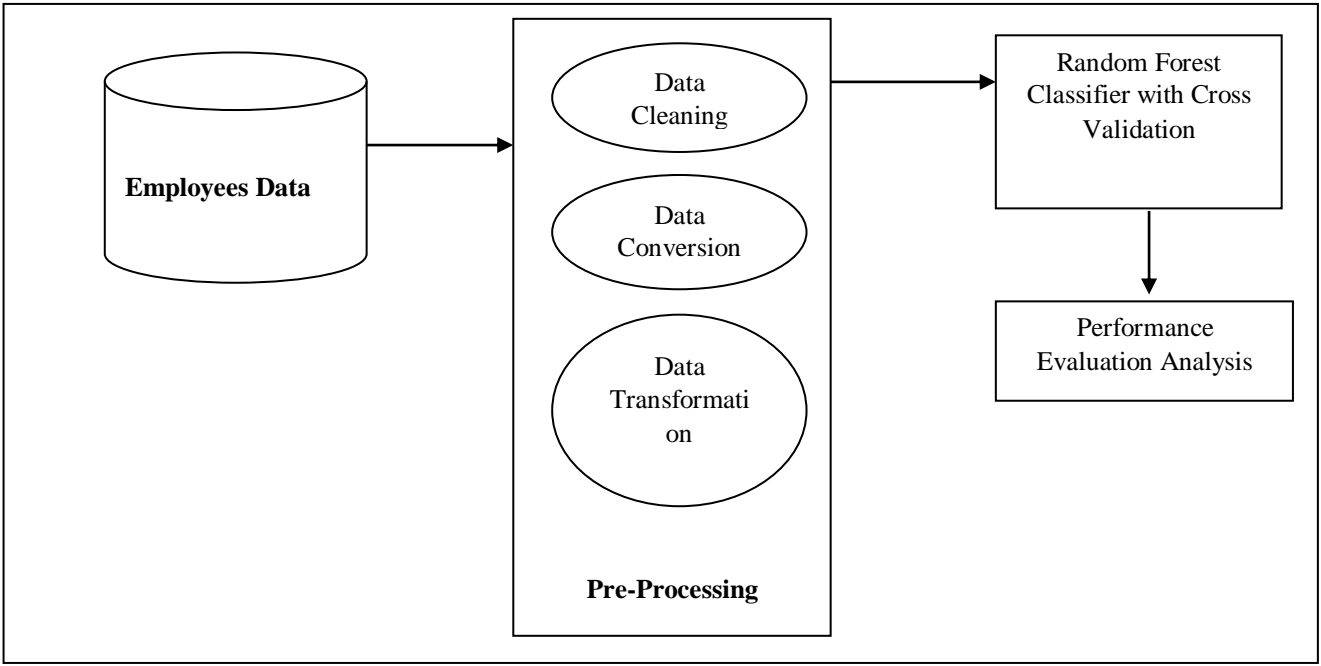

**Figure 3: Block Diagram for HR prediction over Employee Attrition.**

**Algorithm for Random Forest with LOOCV:**

Input: D← Employee Dataset

Output: The accuracy of the model

Begin:

 1. D← apply median strategy to fill the missing numerical data

 2. D← apply mode strategy to fill the missing numerical data

3. for i: 0 to n:

if D[i].dtype=="Object" && D[i].unique_values()<=5:

callLabelEncoder()

    else if D[i].dtype=="Object"

callOnehotEncoder()

4. for i:0 to n:

D[i]←perform absolute max scaler normalization

5.hr_X, hr_y←split the data into training and testing

6. model←RandomForestClassifier(n_estimators)

7. model_cv←LeaveOneOut()

8. result←cross_validation_score(model, hr_X, hr_y,scoring="accuracy",cv_val=model_cv)

 9. print the result

End

Working of Random Forest Classifier: With the advancement of machine learning approaches, ensemble algorithms are getting more attention from the researchers. One type of bagging ensemble algorithm is "Random Forest". The Random Forest constructs a set of forests by combining the decision trees generated. Let us a work on a sample data as shown in the table 4 for seeing how the decision trees are generated.

**Table 4: Sample Dataset**

| S.No | City_deve | relevent_experience | enrolled_university | target |
|------|-----------|---------------------|---------------------|--------|
| 1 | 0.92 | Has relevent experience | no_enrollment | 1 |
| 2 | 0.776 | No relevent experience | no_enrollment | 0 |
| 3 | 0.624 | No relevent experience | Full time course | 0 |
| 4 | 0.789 | No relevent experience | Full time course | 1 |
| 5 | 0.767 | Has relevent experience | no_enrollment | 0 |

| 6 | 0.764 | Has relevent experience | Part time course | 1 |
| 7 | 0.92 | Has relevent experience | no_enrollment | 0 |
| 8 | 0.92 | Has relevent experience | no_enrollment | 1 |

In the sample dataset system has considered three independent variables known as city_deve, relevant_experience, enrolled_university and one dependent variable known as target.

Step 1: The entropy for the target (class label)is calculated as follows:

$Entropy(target) = -[\frac{4}{8}\log_2\frac{4}{8} + \frac{4}{8}\log_2\frac{4}{8}] = -1$

Step 2: Let us calculate the average weighted entropy of enrolled_university by analysing the data as shown in the table 5.

**Table 5: Data Analysis on the attribute "enrolled_university"**

| Feature Value | 0 | 1 | Total |
|---|---|---|---|
| no_enrollment | 3 | 2 | 5 |
| Full time course | 1 | 1 | 2 |
| Part time course | 0 | 1 | 1 |
| | | | **8** |

E(D,enrolled_university)= $\frac{5}{8}$ * E(3,2) + $\frac{2}{8}$ * E(1,1) + $\frac{1}{8}$ * E(0,1)

$= \frac{5}{8}* -[\frac{3}{5}\log_2\frac{3}{5}+\frac{2}{5}\log_2\frac{2}{5}]+ \frac{2}{8} * -[\frac{1}{2}\log_2\frac{1}{2}+\frac{1}{2}\log_2\frac{1}{2}] + \frac{1}{8} * -[\frac{0}{1}\log_2\frac{0}{1}+\frac{1}{1}\log_2\frac{1}{1}]$

$= \frac{5}{8} * -(-0.438-0.528) + \frac{2}{8} * -(-0.5-0.5) + \frac{1}{8} * -(0-0)$

$= \frac{5}{8} * 0.966 + \frac{2}{8} * 1$

$= 0.603 + 0.25 \rightarrow \textbf{0.853}$

Therefore, the entropy of the enrolled_university is **"0.853".**

Step 3: Let us calculate the average weighted entropy of relevent_experience by analysing the data as shown in the table 6.

**Table 6: Data Analysis on the attribute "relevent_experience"**

| Feature Value | 0 | 1 | Total |
|---|---|---|---|
| Has relevent experience | 2 | 3 | 5 |
| No relevent experience | 1 | 2 | 3 |
| | | | **8** |

E(D,relevent_experience)= $\frac{5}{8}$ * E(2,3) + $\frac{3}{8}$ * E(1, 2)

$= \frac{5}{8}* -[\frac{2}{5}\log_2\frac{2}{5}+\frac{3}{5}\log_2\frac{3}{5}]+ \frac{3}{8} * -[\frac{1}{3}\log_2\frac{1}{3}+\frac{2}{3}\log_2\frac{2}{3}]$

$= \frac{5}{8} * -(-0.528-0.438) + \frac{3}{8} * -(-0.524-0.389)$

$= \frac{5}{8} * 0.966 + \frac{3}{8} * 0.913$

$= 0.603 + 0.342 \rightarrow 0.945$

Therefore, the entropy of the relevent_experience is **"0.945".**

Step 4: since the City_deve is the continuous variable, the weighted average of the entropy is calculated based on the mean calculation. The entropy is considered as >= 0.81 and<0.81. Let us calculate the average weighted entropy of City_deve by analysing the data as shown in the table 7

**Table 7: Data Analysis on the attribute "City_deve"**

| Feature Value | 0 | 1 | Total |
|---|---|---|---|
| >=0.81 | 2 | 1 | 3 |
| <0.81 | 3 | 2 | 5 |
| | | | **8** |

E(D, City_deve )= $\frac{3}{8}$ * E(2,1) + $\frac{5}{8}$ * E(3, 2)

$= \frac{3}{8}* -[\frac{2}{3}\log_2\frac{2}{3}+\frac{1}{3}\log_2\frac{1}{3}]+ \frac{5}{8} * -[\frac{3}{5}\log_2\frac{3}{5}+\frac{2}{5}\log_2\frac{2}{5}]$

$= \frac{3}{8} * -(-0.389-0.524) + \frac{5}{8} * -(-0.438-0.528)$

$= \frac{3}{8} * 0.913 + \frac{5}{8} * 0.966$

$= 0.342 + 0.603 \rightarrow 0.945$

Therefore, the entropy of the City_deve is **"0.945".**

Step 5: Now calculate the information gain for all the independent variables as difference of particular attribute entropy and entropy of the class label

Information gain(D,enrolled_course)= 1-0.853 $\rightarrow$ 0.147

Information gain(D, relevent_experience) 1-0.945 $\rightarrow$ 0.005

Information gain(D, city_deve) 1-0.945 $\rightarrow$ 0.005

Step 6: The attribute with highest information gain will be treated as root node, but two attributes have same highest values so based on these two trees can be generated as shown in figure 4

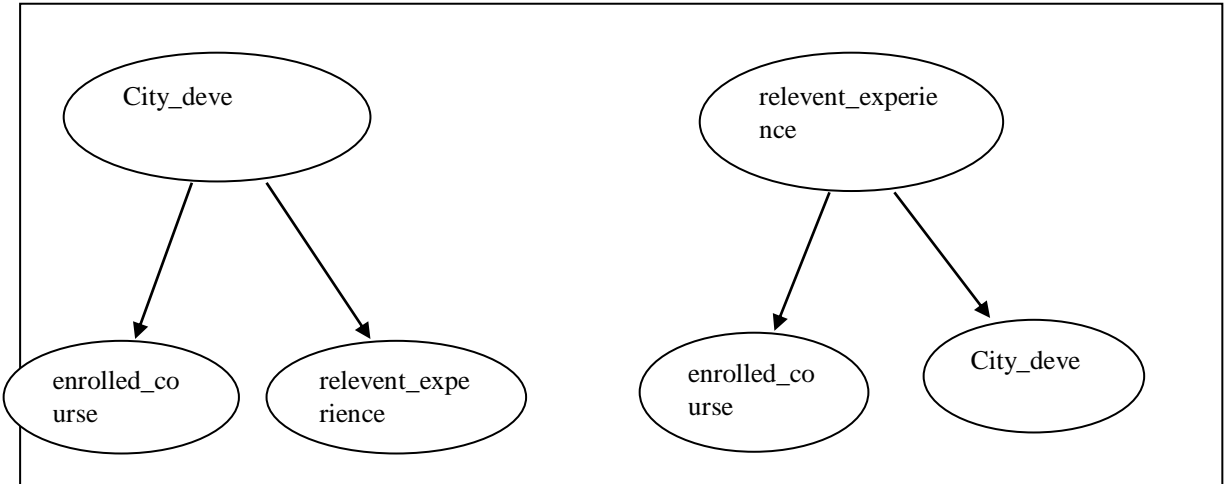

**Figure 4: Generation of Decision Trees**

**Experimental Results:** The proposed system compared the results with traditional machine learning algorithms and represented in **figure 5**. The Logistic Regression (LR) algorithm works on the phenomenon of probability in predicting the class labels. In LR, the output value always either 0 or 1. In Gradient boosting algorithm, the decision depends on the multiple trees, which are weak. The proposed model is compared with two flavours of boosting algorithms out of which one is Extreme Gradient boosting algorithm, which constructs the tree by adjusting the errors obtained in the previous steps. The second one is Light Gradient boosting algorithm, which produces the output at fast rate but it, does not suit for the small amount of data, like the dataset used in this model.

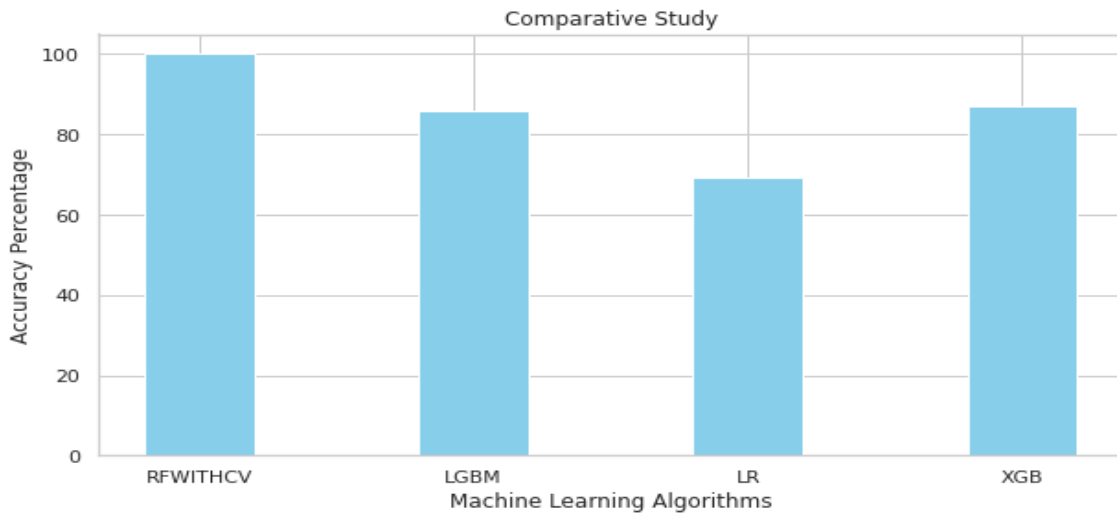

**Figure 5: Comparative Study on Machine Learning Algorithms**

The evaluation metrics recall, accuracy, f1-score and support of the model is represented in the **table 8** .

**Table 8: Evaluation Metrics**

|  | precision | recall | f1-score | support |
|---|---|---|---|---|
|  | 1.00 | 1.00 | 1.00 | 14381 |
|  | 1.00 | 1.00 | 1.00 | 14381 |
| **Accuracy** |  |  | 1.00 | 28762 |
| **Macro avg** | 1.00 | 1.00 | 1.00 | 28762 |
| **Weighted avg** | 1.00 | 1.00 | 1.00 | 28762 |

**Conclusion:** Data science helps the business organizations to take supportive decisions effectively. This model was created to help HR researchers better understand the factors that cause people to leave their current jobs. You can predict whether an applicant will search for a new job or stay with the organisation by using model(s) that use current qualifications, demographics, and experience data, as well as interpret affected factors on employee decision. . Candidates' demographics, schooling, and expertise are all in the hands of those who sign up and enrol. The machine learning algorithms are popular for the prediction process. The proposed model uses random forest classifier and obtained an accuracy of 100%, which can be claimed as "overfitting". This problem generally occurs due to training over the noisy and unwanted data. In the future, studies can design a model that prunes the tree that learns many details.